\documentclass[aps, prl, superscriptaddress, twocolumn, noshowpacs]{revtex4}

\usepackage[utf8]{inputenc}
\usepackage{siunitx}
\usepackage{graphicx}
\usepackage{amsmath,amssymb}
\usepackage{enumitem}
\usepackage{subcaption}
\usepackage{xcolor}
\usepackage{url}
\usepackage{amsmath}
\usepackage{bm}
\usepackage{hyperref}

\begin{document}
\title{A criterion for critical junctions in elastic-plastic adhesive wear}

\author{Stella Brach\footnote{Corresponding author: stella.brach@polytechnique.edu}}
\affiliation{École Polytechnique, 91120 Palaiseau, France}

\author{Sylvain Collet}
\affiliation{École Polytechnique, 91120 Palaiseau, France}

\begin{abstract}
We investigate elastic-plastic adhesive wear via a continuum variational phase-field approach. 
The model seamlessly captures the transition from perfectly brittle, over quasi-brittle to elastic-plastic wear regimes, as the ductility of the contacting material increases. Simulation results highlight the existence of a critical condition that morphological features and material ductility need to satisfy for the adhesive junction to detach a wear debris. We propose a new criterion to discriminate between non-critical and critical asperity contacts, where the former produce negligible wear while the latter lead to significant debris formation.
\end{abstract}

\maketitle

Adhesive wear is one of the most common forms of tribological interaction, whose origins are still incompletely understood  \cite{blau-1997}. It usually occurs in the sliding contact of materials with comparable hardness and in the absence of adequate lubrication \cite{belyi1977adhesive,totten2017friction,bhushan2000modern,stachowiak2013engineering}. 
During the dry sliding of two rough surfaces, interacting asperities are subjected to high contact pressures. This leads to the formation of strong adhesive bonds at the contact interfaces. Upon further sliding, junctions fail by either plastic smoothing or fracture-induced debris formation. 
The first wear mechanism was described by Bowden and Tabor \cite{bowden2001friction}. It is consistent with the Holm's model \cite{holm-1946} and with
 atomic force  microscope experiments \cite{bhaskaran2010ultralow,maw2002single,merkle2008liquid,gotsmann2008atomistic}. 
The second wear behavior was predicted by Archard \cite{archard-1953} and reported in experiments at different length-scales 
 \cite{brockley-1965, green1955friction,liu2010method, sato2012real}.

Predicting the wear behavior of a single-asperity contact holds great promise for technological innovation.
It can steer the development of the next generation archival storage \cite{millipede1,millipede2} and scanning probes  \cite{dai1998exploiting,lantz2012wear}. Transposing this knowledge to rough surfaces can lead to the design of tribosystems with reduced wear damage and of tailor-made lubricants \cite{holmberg-2017}. 
Despite the substantial research effort, predicting the onset of wear still is a major challenge \cite{blau-1997}. This resulted in the emergence of a variety of modeling approaches aiming, among other objectives, at reproducing the above-mentioned interplay between plastic smoothing and debris formation \cite{vakis-2018}.

The transition between these two wear mechanisms was captured for the first time in \cite{aghababaei-2016}, by using a discrete description of a model single-asperity contact.
Simulation results revealed the existence of a \textit{critical length-scale}, which controls the adhesive wear process: smaller junctions are gradually smoothed out during sliding, while bigger ones detach a wear debris.
To achieve this result, it was nonetheless necessary to sensibly embrittle the material by using a coarse-grained potential with an artificially reduced size of the crack tip plastic zone  \cite{aghababaei-2016, milanese2019emergence}. 
On the other hand, most of continuum approaches focus on the definition of plastic contact \cite{pei2005finite,jacq2002development,frerot2019fourier} and real contact area \cite{persson2001elastoplastic,majumdar-1990}, while very few works address the removal of wear debris \cite{akchurin2016stress,li2019mesoscale,frerot2020crack,popov2018adhesive}.

In this Letter, we investigate adhesive wear in elastic-plastic junctions by means of a variational phase-field approach \cite{bourdin-2008,brach2019phase}. This enables us to address a wide range of material behaviors and junction geometries, at a relatively low computational cost. With respect to discrete approaches, the proposed model is \textit{seamless}. It requires no artificial adjustment of the constitutive laws of the material when going from brittle to ductile behaviors, and naturally captures the transition between fracture-dominated and plasticity-dominated wear mechanisms as ductility increases. A new criterion is proposed to discriminate between \textit{non-critical} and \textit{critical} adhesive junctions, where the former fail with no or negligible mass loss while the latter lead to significant debris formation.

Adhesive wear is understood as a fracture problem \cite{Collet-2020,carollo-2019}. This is solved by using a variational phase-field approach, coupled with a rate-independent plasticity formulation \cite{brach2019phase}. 
 Sharp cracks are regularized by introducing a scalar phase (or damage) field.
A regularized energy functional is minimized with respect to the plastic strain, damage and displacement field, as detailed in the Supplemental Material SM.
The contacting materials are linear elastic and perfectly plastic, with von Mises yield strength $\sigma_\text{y}$ and ultimate strength $\sigma_\text{uts}$.
The ratio $r_\text{y}=\sigma_\text{uts}/\sigma_\text{y}$ discriminates between quasi-brittle ($r_\text{y}<1$) and ductile ($r_\text{y}>1$) fracture \cite{brach2019phase,brach2020effects}.
The case of a brittle material is recovered for $r_\text{y}\to 0$.

\begin{figure}[t!]
    \centering
    \includegraphics[width=0.48\textwidth]{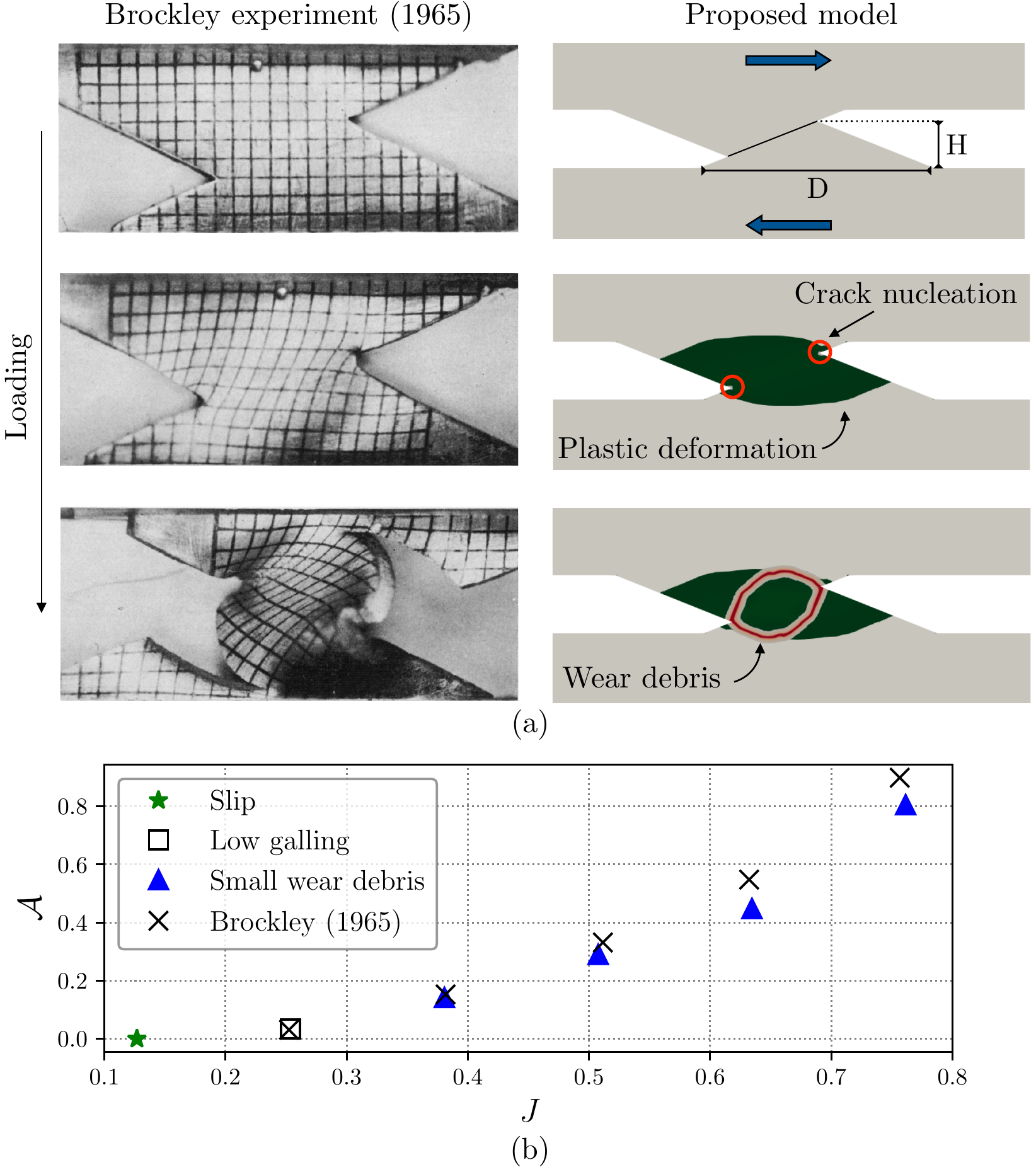}
    \caption{{
    Assessment of the proposed model against experiments \cite{brockley-1965}. $H/D=0.28$, $r_\text{y}=3$.
    (a) Wear process for $J=0.76$. (b) Wear damage $\mathcal{A}$ vs. junction length $J$.
    }}
    \label{fig:brockley}
\end{figure}

We consider the experiment proposed by Green \cite{green1955friction} and Brockley \cite{brockley-1965},
where two triangular asperities of base $D$ and height $H$ form an adhesive junction (Fig.\ref{fig:brockley}). 
The non-dimensional junction length $J\in[0,1]$ is defined as the ratio between the contact interface and the asperity side, such that $J=0$ (resp., $J=1$) corresponds to a tip-to-tip (resp., tip-to-base) contact.
A quasi-static shear loading is imposed by the relative horizontal displacement of the top and bottom boundaries, while the vertical displacement is fixed to zero. 
For purely brittle materials, the criterion for debris formation is \cite{Collet-2020}
\begin{equation}
    \label{eq:brittle_cr}
    J \geq \,\mathcal{C}\,\frac{D}{H}\,,
\end{equation}
 which corresponds to a Griffith energy balance between the  elastic  energy  stored  in  the  junction  and  the fracture energy necessary to detach a debris (see SM). In \eqref{eq:brittle_cr}, $\mathcal{C}$ is a non-dimensional constant that  describes the minimum slenderness for which fully engaged asperities detach a large debris.
If condition \eqref{eq:brittle_cr} is satisfied, a wear particle detaches from the sliding contact.
The junction is then called \textit{critical}.
Otherwise, no or negligible mass loss is observed. 
The curve $J^\ast=\mathcal{C}\,\cdot D/H$ identifies the transition between the two wear behaviors, for a junction of given slenderness $H/D$ and critical length $J^\ast$ (Fig.\,1 SM).

The observed wear mechanisms are classified as (Fig.\,2 SM): \textit{slip} (asperities slide past one another), \textit{galling} (transfer of material from one asperity to the other) and \textit{particle formation} (detachment of a wear debris).
The ratio $\mathcal{A}$ between the worn area and the undamaged asperity surface (i.e., $HD/2$) is used to determine the severity of the wear damage,  as detailed in Eq.\,(4) of the SM.

We begin by validating our numerical approach against the experimental results provided by Brockley and Fleming \cite{brockley-1965}. 
 The original setup is recovered by setting $H/D$ and $J$ equal to the values reported in \cite{brockley-1965}.
Model parameters are chosen to reproduce the behavior of the B152 ASTM copper  used in \cite{brockley-1965}. Details are provided in the SM.
Figure \ref{fig:brockley}\,(a) shows the comparison between the computed wear response and the experiment.
Upon sliding, the junction plastically deforms.  
Tensile stresses concentrate at the edges of the adhesive interface and trigger the nucleation of a pair of cracks. The  junction eventually fails by detaching a small wear debris, as observed in the experiment.
 In Fig.\,\ref{fig:brockley}\,(b), shear tests are performed on junctions with different overlap between the asperities.
The wear damage $\mathcal{A}$ increases as $J$ increases, while the adhesive junction fails through different wear mechanisms. A transition between slip, galling and small particle formation is observed, as a larger portion of the junction plastically deforms. The obtained results are in very good agreement with the experimental data. Discrepancies with the values of $\mathcal{A}$ in \cite{brockley-1965} might be due to the large deformations observed in the experiment as $J$ increases, which are not captured by the present model.

  \begin{figure}[t!]
    \centering
    \includegraphics[width=0.48\textwidth]{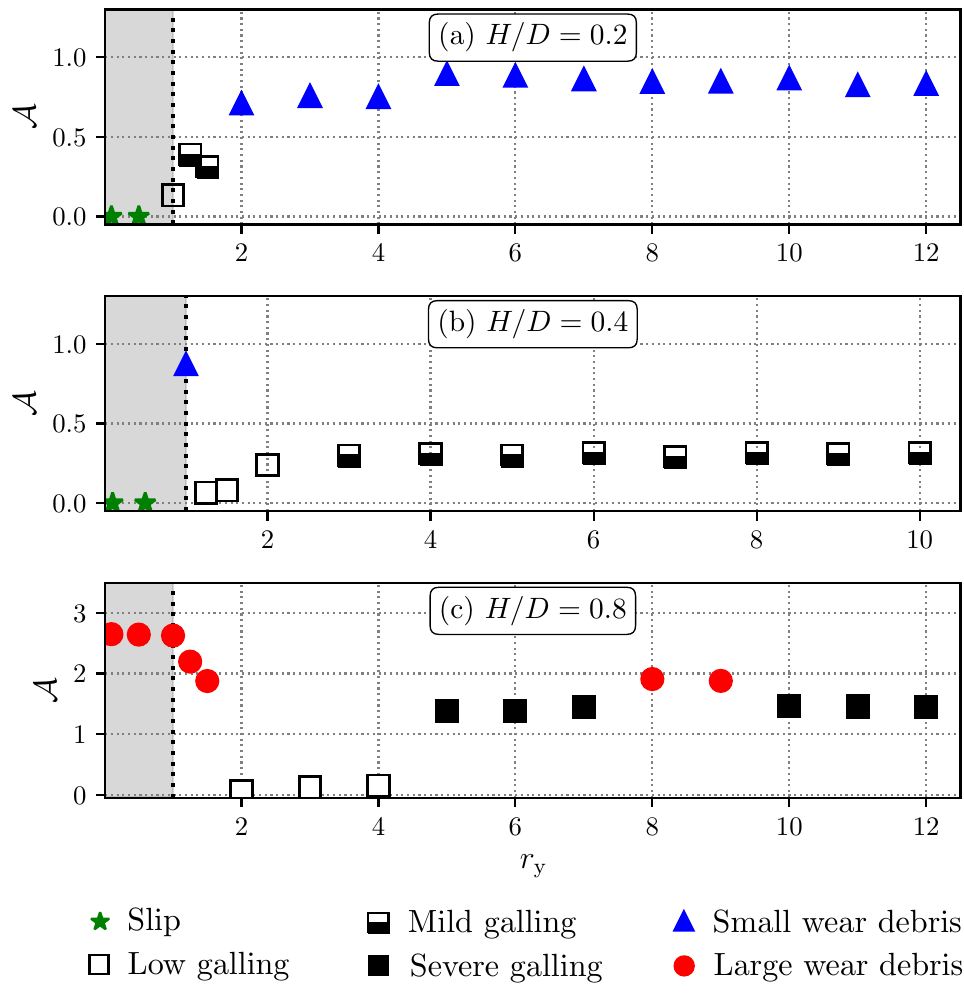}
    \caption{{Wear damage $\mathcal{A}$ vs. ductility ratio $r_\text{y}$. Wear mechanisms observed for quasi-brittle ($r_\text{y}<1$) and ductile ($r_\text{y}>1$) adhesive junctions. $J=0.7$.
    }}
    \label{fig:plasticity}
\end{figure}

  \begin{figure}[bt!]
    \centering
    \includegraphics[width=0.48\textwidth]{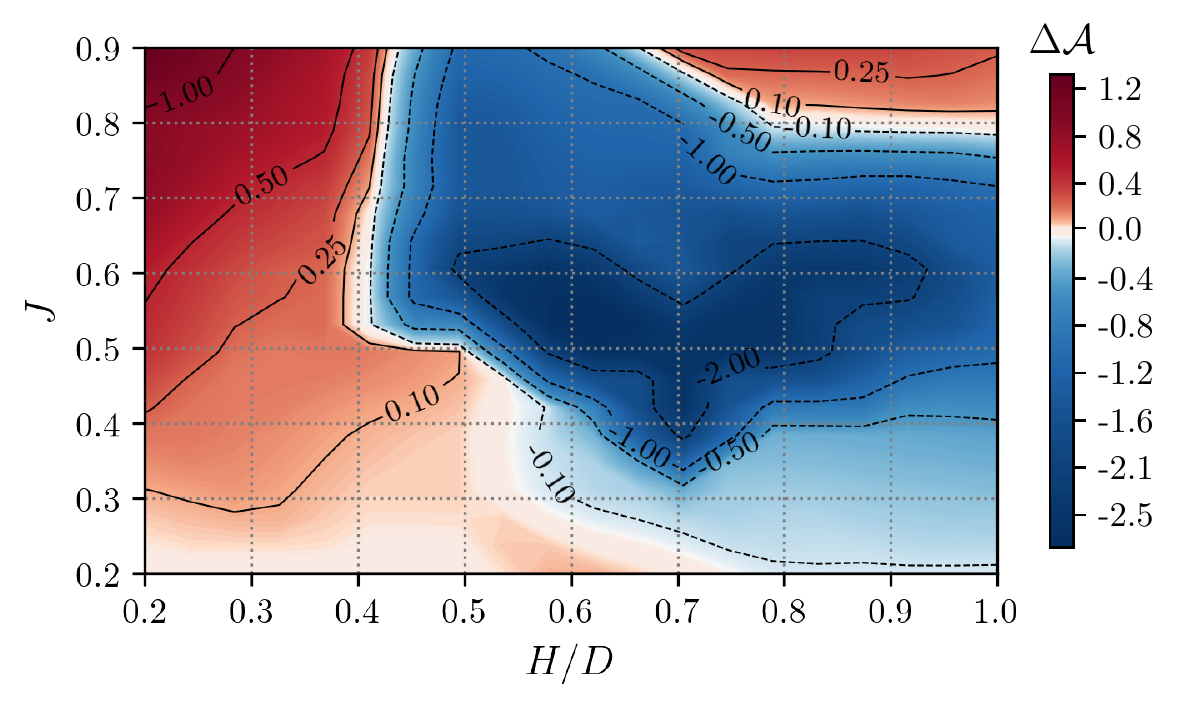}
    \caption{{Difference in wear damage $\Delta\mathcal{A}$ between ductile ($r_\text{y}=6$) and quasi-brittle ($r_\text{y}=0.5$) junctions.}}
    \label{fig:archard}
\end{figure}

  \begin{figure*}[bt!]
    \centering
    \includegraphics[width=1\textwidth]{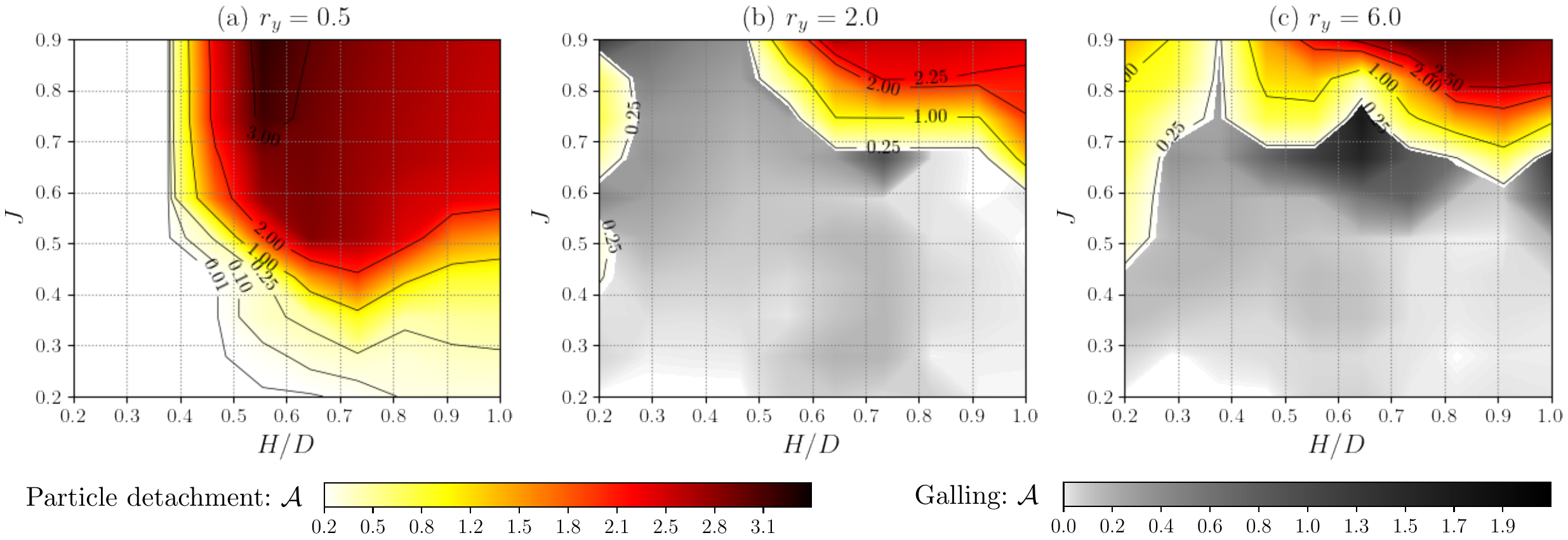}
    \caption{{
    Evolution of the wear damage $\mathcal{A}$, for different ductility ratios $r_\text{y}$. Galling and particle detachment are highlighted. Contour plots display the severity of the wear damage caused by debris formation.}}
    \label{fig:maps}
\end{figure*}

The influence of plasticity on the wear response of the adhesive junction is analysed in Fig.\,\ref{fig:plasticity}, for a given $J$ and increasing values of $H/D$. The ductility ratio $r_\text{y}$ is varied to consider quasi-brittle ($r_\text{y}<1$) and elastic-plastic ($r_\text{y}>1$) contacts.
In Figs.\,\ref{fig:plasticity}\,(a)-(b), the junction's geometry does not satisfy the criterion \eqref{eq:brittle_cr}. In fact, quasi-brittle junctions fail through slip and do not form any wear debris.
For $r_\text{y}>1$, the region surrounding the adhesive interface plastically deforms to accommodate the large contact stresses. 
For $H/D=0.2$, we observe a transition from slip to galling, and from galling to particle detachment as $r_\text{y}$ increases.
Conversely, for $H/D=0.4$, galling is favoured over debris formation as ductility increases. 
Hence, for a given choice of material (i.e., $r_\text{y}$) and contact conditions (i.e., $J$), a small variation in the asperity slenderness can lead to completely different wear responses. This softens the conclusion reported in \cite{brockley-1965}, according to which \textit{``all junctions of the type investigated will give rise to a wear particle''}. 
In Fig.\,\ref{fig:plasticity}\,(c), $H/D$ and $J$ satisfy the criterion \eqref{eq:brittle_cr}. Quasi-brittle junctions thus fail by detaching a large wear debris.  As $r_\text{y}$ increases, particle detachment is replaced by low galling, and part of the material comprising the junction is transferred from one surface to the other. This transition is followed by severe galling in case of more ductile contacts.
Small perturbations of the wear behavior are still observed for some values of $r_\text{y}$, which reveals a strong competition between fracture-induced and plasticity-dominated wear. This is shown in Fig.\,3 of the SM. Depending on the local stress distribution, it can happen that a secondary crack joins the main fracture process, producing a wear debris (Fig.\,3\,(a) SM). In most of the cases though, secondary cracks -if present- do not contribute to the failure of the junction, and galling is eventually observed (Fig.\,3\,(b) SM).
In all the elastic-plastic cases in Fig.\,\ref{fig:plasticity}, $\mathcal{A}$ increases with $r_\text{y}$, until converging to a plateau value. At this point, the whole junction is plastically deformed and further increase of ductility only results in a further extension of the plastic zone into the bulk surfaces.

According to the Archard's law \cite{archard-1953}, the harder the junction, the smaller is the wear.  
Some authors however object that a more complex dependency of the wear rate on the hardness might be dissimulated by the large variability of  Archard's wear coefficient \cite{popov2018adhesive,popov2019generalized}.
Our results show that morphological features also play a fundamental role.
In Figs.\,\ref{fig:plasticity}\,(a)-(b),  $\mathcal{A}$ increases with $r_\text{y}$, consistently with the Archard's law. The model also captures the dependency of galling on ductility, as weak materials gall more than hard ones. 
An opposite trend is observed in Fig.\,\ref{fig:plasticity}\,(c), where $\mathcal{A}$ initially reduces with increasing $r_\text{y}$. 
Archard's proportionality is recovered after the change in wear mechanism, as galling becomes more severe with $r_\text{y}$. The plateau value reached by $\mathcal{A}$ is however lower than that experienced in the brittle regime.

Figure \ref{fig:archard} further investigates the dependency of wear damage on hardness, by comparing the values of $\mathcal{A}$ obtained for quasi-brittle ($r_\text{y}=0.5$) and ductile ($r_\text{y}=6$) junctions. The ductility ratio is chosen to capture the two plateaux in Fig.\,\ref{fig:plasticity}. Variations $\Delta\mathcal{A}=\mathcal{A}|_{r_\text{y}=6}-\mathcal{A}|_{r_\text{y}=0.5}$ are represented in the plane defined by the junction length $J$ and the asperity slenderness $H/D$.
Two distinct trends clearly arise. On one hand, ductile junctions with small $H/D$ wear more than their quasi-brittle counterparts. This is in agreement with both the Archard's law and the physical insights provided in \cite{frerot2020crack}.
This increase in wear damage is particularly significant for large values of $J$, where a part of the adhesive junction is debonded during sliding (Figs.\,\ref{fig:plasticity}\,(a)-(b)). 
Conversely, ductile junctions with larger $H/D$ wear less than in the brittle regime, mostly due to the occurrence of galling. 
The wear behavior of those junctions does not follow Archard's prediction, when comparing the two plateaux at $r_\text{y}=0.5$ and $r_\text{y}=6$. Note however that after transitioning to the ductile regime, most of those junctions follow the evolution shown in Fig.\,\ref{fig:plasticity}\,(c), and thus gall more with increasing ductility.
Finally, Fig.\,\ref{fig:archard} shows that Archard's proportionality is recovered for large values of both $H/D$ and $J$, for which $\Delta\mathcal{A}$ becomes again positive.
Importantly, the comparison in Fig.\,\ref{fig:archard} only applies to early-stage wear. 
Both the sign and the intensity of $\Delta\mathcal{A}$ are expected to evolve during sliding, as galling and third-body interactions progressively modify the surface roughness.

Figure \ref{fig:plasticity} demonstrates that the criterion \eqref{eq:brittle_cr} only holds for brittle and quasi-brittle adhesive junctions. Its possible extension to elastic-plastic contacts is investigated in Fig.\,\ref{fig:maps}. 
Each map reports the wear mechanisms observed for a given $r_\text{y}$ and the corresponding wear damage $\mathcal{A}$. 
To enhance readability, only galling and particle formation are highlighted, while the remaining junctions fail through slip. We find that the wear response strongly depends on ductility. For quasi-brittle contacts in Fig.\,\ref{fig:maps}\,(a), the left side of the map is occupied by wide asperities that do not produce any wear debris. 
As the material becomes more ductile, an increasing number of those junctions fail by detaching a small particle. The setup considered in \cite{brockley-1965} belongs to this region. We observe a transition from slip to galling and from galling to particle detachment, when going from Fig.\,\ref{fig:maps}\,(a) to Fig.\,\ref{fig:maps}\,(c).
On the other hand, the right side of Fig.\,\ref{fig:maps}\,(a) is occupied by brittle junctions complying with Eq.\,\eqref{eq:brittle_cr}, which detach large wear debris. 
With increasing $r_\text{y}$, the extent of this region progressively reduces, while particle detachment is mostly replaced by galling.
For the most ductile material in Fig.\,\ref{fig:maps}\,(c), only junctions belonging to the top-right corner of the map produce large debris. The most severe forms of galling occur at the boundaries of this region.
Those results suggest that there must exist a critical condition that $H/D$, $J$ and $r_\text{y}$ need to satisfy for the sliding contact to fail through one wear mechanism or the other.

In \cite{aghababaei-2016}, an adhesive junction is defined \textit{critical} if it fails by ``\textit{debonding the two asperities from the sliding surfaces}''. This corresponds to particle detachment with wear damage $\mathcal{A}\geq 2$. 
We use the phase-field model to compute wear maps as in Fig.\,\ref{fig:maps}, for materials with different levels of ductility. 
We then apply the above condition to identify critical junctions, by neglecting isolated perturbations of the wear response as those observed in Fig.\,\ref{fig:plasticity}\,(c). 
We find that all the asperity contacts satisfying
\begin{equation}
    \label{eq:elpl_cr}
    J \geq \gamma \,\left(\frac{D}{H}\right)^\delta\,,
\end{equation}
with $\gamma=\gamma(r_\text{y})$ and $\delta=\delta(r_\text{y})$, debond the entire junction during sliding. Geometries that do not comply with \eqref{eq:elpl_cr} fail through slip, galling, or detach particles by causing limited wear damage ($\mathcal{A}<2$). 

  \begin{figure}[b!]
    \centering
   {\includegraphics[width=0.48\textwidth]{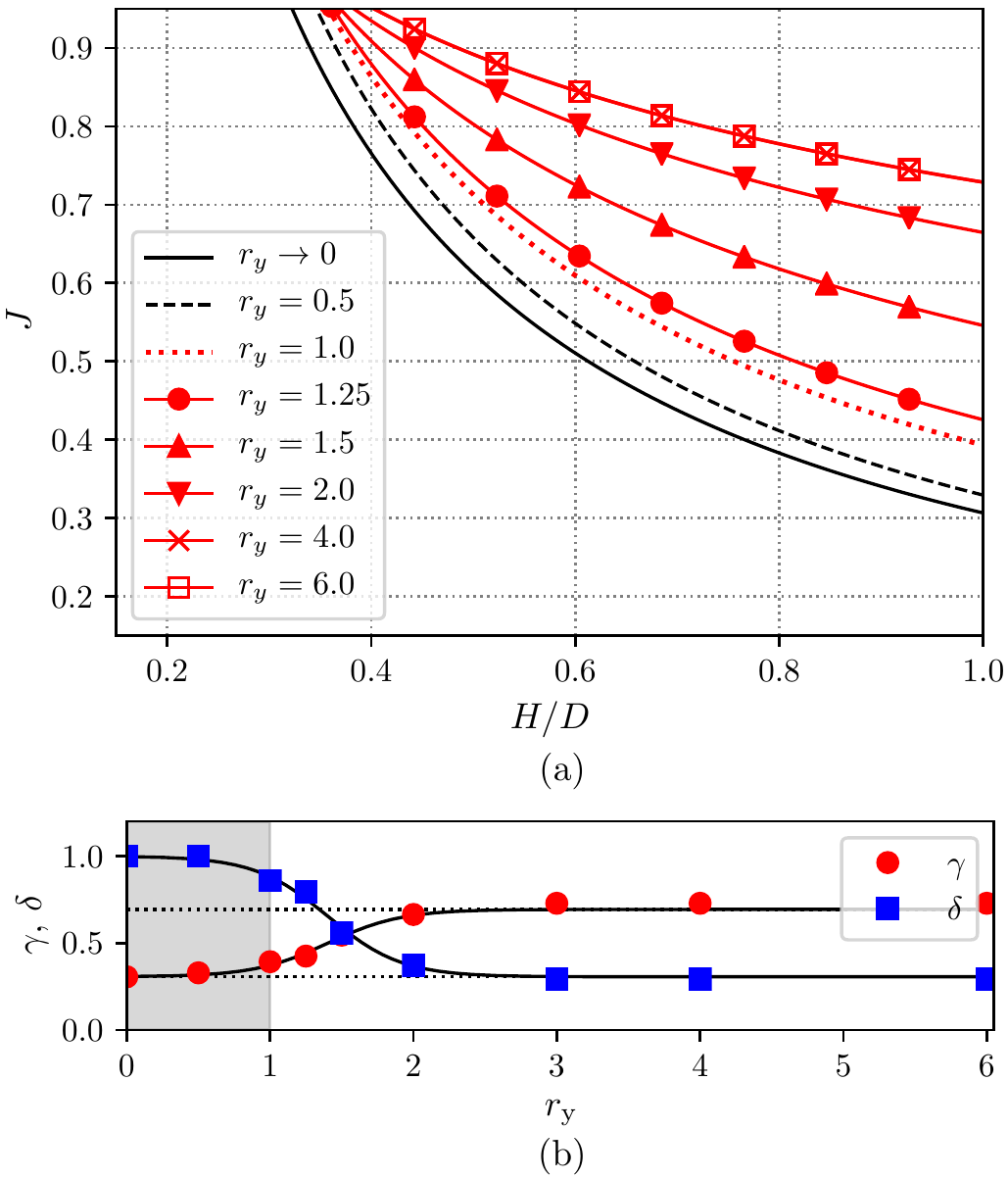}}
    \caption{{
    (a) Criterion \eqref{eq:elpl_cr}.
    (b) Coefficients $\gamma$ and $\delta$: computed values (symbols) and predictions \eqref{eq:param} with 
    $s=1.4$, $w=0.25$ and $\mathcal{C}=0.3$. Dotted lines in (b) mark the plateau values at $\mathcal{C}$ and $1-\mathcal{C}$.}
    }
    \label{fig:critical}
\end{figure}

The inequality constraint \eqref{eq:elpl_cr} is the
criterion that governs adhesive wear in elastic-plastic contacts, by discriminating between \textit{non-critical} (no or negligible mass loss) and \textit{critical} (significant debris formation) junctions. 
Its evolution as a function of the ductility ratio $r_\text{y}$ is shown in Fig.\,\ref{fig:critical} (a), where \eqref{eq:elpl_cr} is represented with the equal sign. The resulting curve delimits the portion of the plane, which is occupied by critical junctions, from below.
 The corresponding values of $\gamma$ and $\delta$ are computed from the wear maps by setting the threshold
 $\mathcal{A}\geq 2$ on debris formation, and are represented with markers in Fig.\,\ref{fig:critical}\,(b).

We observe a smooth transition between two wear regimes: \textit{brittle} ($r_\text{y}\to 0$) and \textit{ductile} ($r_\text{y}\gg 1$) adhesive wear.
This transition is well captured by 
\begin{equation}
    \label{eq:param}
    \begin{aligned}
    \gamma &= \mathcal{C}+\left(1-2\mathcal{C}\right)(1-S)\\
    \delta &= 1 - \left(1-\mathcal{C} \right)(1-S)
    \end{aligned}
\end{equation}
where $\mathcal{C}$ is the constant in Eq.\,\eqref{eq:brittle_cr} and $S=S(r_\text{y})$ is the sigmoid function 
$S = \left[1+\exp{\left((r_\text{y}-s)/w\right)}\right]^{-1}$.
The rate of the transition is controlled by $w$, while $s$ is the switch-point that discriminates between the two wear regimes.
Estimates of $w$ and $s$ are obtained by fitting expressions \eqref{eq:param} to the values of $\gamma$ and $\delta$ computed from the wear maps. Solid lines in Fig.\,\ref{fig:plasticity}\,(b) represent the resulting evolution curves as a function of the ductility ratio.
 For brittle materials, $\delta=1$ and the criterion \eqref{eq:elpl_cr} recovers the original condition \eqref{eq:brittle_cr} with $\mathcal{\gamma}=\mathcal{C}$. 
In case of elastic-plastic contacts, $\delta$ smoothly decreases with $r_\text{y}$ until reaching a plateau value at $\delta=\mathcal{C}$. The opposite behavior is observed for $\gamma$, which increases with ductility up to the constant value $\gamma=1-\mathcal{C}$.
As a result, in Fig.\,\ref{fig:critical} (a), the extent of the region occupied by critical junctions reduces with increasing $r_\text{y}$, and eventually becomes constant for the most ductile materials. 
This evolution accommodates the transition from particle detachment to increasingly severe forms of galling, as shown in Figs.\,\ref{fig:plasticity}-\ref{fig:maps}.

In summary, we proposed a continuum phase-field model that seamlessly captures the transition from brittle to ductile adhesive wear.
Simulation results highlighted the existence of a critical condition that the contact geometry and the ductility of the material need to satisfy for the junction to produce a wear debris. The resulting criterion enables wear prediction. It can lead to development of algorithms for roughness evolution \cite{frerot-2018}, and to the design of surfaces with reduced wear damage \cite{kim2010tribology}.

\vspace{1em}
\begin{acknowledgments}
We are grateful to Dr.\,B. Dorschner for the fruitful discussions  and to the anonymous Reviewers for the pertinent comments.
Computational resources at the GENCI-IDRIS center were under the Grant 2020-[A0090911874].
\end{acknowledgments}

\bibliographystyle{apsrev4-1}
%\bibliography{biblio.bib}

%

\end{document}